\documentclass[12pt]{iopart}
\usepackage[]{graphicx}
\usepackage[dvipdfm,colorlinks,linkcolor=blue,citecolor=blue]{hyperref}
\usepackage{amssymb}

\newcommand{\sss}[1]{\scriptscriptstyle{#1}}
\newcommand{\gf}[2]{\langle\!\langle #1 ; #2 \rangle\!\rangle}

\newcommand{\ave}[1]{\langle #1 \rangle}
\newcommand{\avez}[1]{ {\langle #1 \rangle}_{\!\sss{0}}}

\newcommand{\mySp}[2]{\hat{S'}\vphantom{\hat{S}}^{\phantom{\prime}}_{\!\!2\mybf{#1}}{\!\!\!^{#2}}}
\newcommand{\onlinecite}[1]{{\cite{#1}}}
\newcommand{\dfrac}[2]{{\displaystyle \frac{#1}{#2}}}
\newcommand{\text}[1]{{\rm #1}}
\newcommand{\bm}[1]{{\bi{#1}}}
\newcommand{\TN}{T_\text{N}}

\begin{document}

\title[Gui-Bin Liu and Bang-Gui Liu]
  {Temperature-dependent striped antiferromagnetism of LaFeAsO in a Green's function approach}

\author{Gui-Bin Liu and Bang-Gui Liu}

\address{Institute of Physics, Chinese Academy of Sciences, Beijing 100190, China\\
Beijing National Laboratory for Condensed Matter Physics, Beijing 100190, China}

\ead{gbliu@aphy.iphy.ac.cn {\rm and} bgliu@aphy.iphy.ac.cn}
\date{\today}

\begin{abstract}
We use a Green's function method to study the
temperature-dependent average moment and magnetic phase-transition
temperature of the striped antiferromagnetism of LaFeAsO and other
similar compounds as the parents of FeAs-based superconductors. We
consider the nearest and the next-nearest couplings in the FeAs
layer and the nearest coupling for inter-layer spin interaction.
The dependence of the transition temperature $\TN$ and the
zero-temperature average spin on the interaction constants are
investigated. We obtain an analytical expression for $\TN$ and
determine our temperature-dependent average spin from zero
temperature to $\TN$ in terms of unified self-consistent
equations. For LaFeAsO, we obtain a reasonable estimation of the
coupling interactions with experimental transition
temperature $\TN=138$ K. Our results also show that a non-zero
antiferromagnetic (AFM) inter-layer coupling is essential to the
existence of a non-zero $\TN$ and the many-body AFM fluctuations
reduce substantially the low-temperature magnetic moment per Fe
towards the experimental value. Our Green's function approach can be
used to other FeAs-based parent compounds and these results should
be useful to understand the physical properties of FeAs-based
superconductors.
\end{abstract}

\submitto{\JPCM} 
\maketitle

\section{\label{sec.intro}Introduction}

The discovery of high temperature superconductor
LaF$_x$FeAsO$_{1-x}$ by Kamihara \textit{et~al}\cite{jacs130_3296}
has triggered world-wide researches on all
aspects of FeAs-based pnictides superconductors and their parent
compounds, namely  $Ln$FeAsO ($Ln$=La\cite{jacs130_3296,nat453_899,prl101_077005,550K,Yildirim,prl101_126401,nematic},
Ce\cite{Ce1,Ce2}, Pr\cite{Pr1}, Nd\cite{Nd1,Nd2}, Sm\cite{Sm1,Sm2,Sm3}, \dots)
and $A$Fe$_2$As$_2$ ($A$=Ca\cite{Ca1,Ca2,Ca3},
Sr\cite{sw-SrFe2As2,Sr2,Ba3Sr3}, Ba\cite{Ba1,Ba2,Ba3Sr3}).
$Ln$FeAsO and $A$Fe$_2$As$_2$ own some common characteristics: a)
they are all of layered structures and have structure transitions
from high-temperature tetragonal to low-temperature orthorhombic
symmetry; b) they all have stripe-like  antiferromagnetic (AFM)
order formed by Fe atoms and the AFM transition temperatures
$\TN$'s are not higher than the structure transition temperatures
$T_\text{S}$'s; c) the onset of superconductivity competes with
the AFM and structure transitions\cite{compt1,compt2}. And they
differ in the sense that $\TN<T_\text{S}$ for $Ln$FeAsO and
$\TN=T_\text{S}$ for $A$Fe$_2$As$_2$. Both of the two series can
be made superconducting by doping them with appropriate dopants or
applying pressures. It should help understand the
superconductivity of FeAs-based materials to elucidate the
corresponding antiferromagnetism of the parent compounds.

LaFeAsO is the prototype and the representative of the parent
compounds of FeAs-based superconductors and thus we focus on the
striped AFM order of undoped LaFeAsO, whose $\TN$ is 138K and
$T_\text{S}$ 156K\cite{nat453_899,prl101_077005}. First-principles
results confirm that the stripe-AFM order is the magnetic ground
state\cite{Yildirim}. Spin-wave approaches were adopted
to give the low-temperature excitation spectra\cite{nematic,lsw,sw-SrFe2As2},
 and the spin-orbit interaction
and $p$-$d$ hybridization are used to understand the observed
small magnetic moment $0.25\mu_\text{B}\sim0.36\mu_\text{B}$ per
Fe at low
temperature\cite{nat453_899,prl101_077005,prl101_126401}. However,
it is highly desirable to describe the magnetic moment from zero
temperature to $\TN$ within a unified theory.

In this paper, we use a Green's function method to study the
temperature-dependent average moment and phase-transition
temperature of the striped antiferromagnetism of LaFeAsO and other
similar compounds as the parents of FeAs-based superconductors. We
consider the nearest and the next-nearest couplings in the FeAs
layer and only the nearest one for the inter-layer spin
interaction. The dependence of the transition temperature $\TN$
and the zero-temperature average spin on the four interactions are
investigated. We obtain an analytical expression for $\TN$ and
determine our temperature-dependent average spin from zero
temperature to $\TN$ in terms of unified self-consistent
equations. For LaFeAsO, we obtain a reasonable estimation of the
coupling interactions with experimental phase-transition
temperature $\TN=138$ K. Our results also show that a non-zero
antiferromagnetic inter-layer coupling is essential to the
existence of a non-zero $\TN$ and the many-body AFM fluctuations
reduce substantially the low-temperature magnetic moment per Fe
towards the experimental value. More detailed results will be
presented in the following.

The remaining part of this paper is organized as follows. In next
section, we shall give our spin model, the Green's function
derivation and our main analytical results. 
In section \ref{sec.result}, we shall present our numerical
results and make corresponding discussions. And 
our conclusion is given in \sref{sec.con}.

\section{\label{sec.model}Effective Model, Green's function derivation, and main analytical results}

\begin{figure}[!htbp]
\begin{center}
\includegraphics[width=5cm]{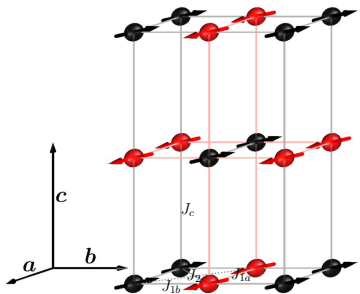}
\end{center}
\caption{The magnetic cell with volume $a\times 2b
\times 2c$ of the orthorhombic Fe spin lattice.  The Fe lattice
consists of two spin sublattices: SL1 (black) and SL2 (red or
gray). $\bm{a}$ and $\bm{b}$ are the two base vectors in the
FeAs layer, and $\bm{c}$ is perpendicular to both of $\bm{a}$ and
$\bm{b}$.} \label{fig:struct}
\end{figure}

To deal with the striped AFM configuration of LaFeAsO, we consider
the Fe lattice of the original orthorhombic LaFeAsO structure and
divide it into two sublattices in each of which the Fe spins align
parallel but antiparallel between the two sublattices
(\fref{fig:struct}). Hence we consider an anisotropic Heisenberg Hamiltonian
\newcommand{\mybf}[1]{\mathbf{#1}}
\newcommand{\mybfa}[2]{\mathbf{#1} + \mathbf{#2}}
\begin{equation}
\hat{H}=\sum_{\ave{\mybf{i,j}}} J_{\ave{\mybf{i,j}}} \mybf{\hat{S}_i} \cdot \mybf{\hat{S}_j} +
        J_2 \sum_{\ave{\!\ave{\mybf{i,j}}\!}} \mybf{\hat{S}_i} \cdot \mybf{\hat{S}_j}~,
\label{eq.H}
\end{equation}
in which $\mybf{\hat{S}}_{\mybf{i}}$ denotes the quantum spin operator
at the lattice position $\mybf{i}$, $\ave{\mybf{i,j}}$ means nearest-neighbour (NN) spin pairs,
and $\ave{\!\ave{\mybf{i,j}}\!}$ means next-nearest-neighbour (NNN) spin pairs in $a$-$b$ plane
(we only consider NNN pairs
in $a$-$b$ plane because the inter-layer interactions are very weak).
NN interaction $J_{\ave{\mybf{i,j}}}$ can be three values:
$J_{1a}$ which is the spin interaction between parallel NN spins in $a$-$b$ plane,
$J_{1b}$ between antiparallel NN spins in $a$-$b$ plane and $J_c$
between inter-layer NN spins. $J_2$ is interaction between NNN spins in $a$-$b$ plane.
Four different $J$'s make $\hat{H}$ anisotropic.
To differentiate spin operators in SL1 and SL2, we use $\mybf{\hat{S}}_{1\mybf{i}}$
and $\mybf{\hat{S}}_{2\mybf{j}}$ to represent them respectively.
For spins in SL2, we make transformations: $\mySp{j}{z}=-\hat{S}_{2\mybf{j}}^z$,
 $\mySp{j}{+}=\hat{S}_{2\mybf{j}}^-$, $\mySp{j}{-}=\hat{S}_{2\mybf{j}}^+$ and then have
\begin{equation}
\eqalign{
\mybf{\hat{S}}_{1\mybf{i}}\cdot\mybf{\hat{S}}_{2\mybf{j}}
            &=\dfrac12(\hat{S}_{1\mybf{i}}^+\hat{S}_{2\mybf{j}}^-+\hat{S}_{1\mybf{i}}^-\hat{S}_{2\mybf{j}}^+)
               +\hat{S}_{1\mybf{i}}^z\hat{S}_{2\mybf{j}}^z\\
            &=\dfrac12(\hat{S}_{1\mybf{i}}^+\mySp{j}{+}+\hat{S}_{1\mybf{i}}^-\mySp{j}{-})
               -\hat{S}_{1\mybf{i}}^z\mySp{j}{z} ~.
}
\label{eq.SS}
\end{equation}
Inserting \eref{eq.SS} into \eref{eq.H} we can get the Hamiltonian expressed by
$\mybf{S}_{1\mybf{i}}$ and $\mybf{S}'_{2\mybf{j}}$ (simple but too long to give out here).

Accordingly, we use Green's function method\cite{gf,Tyablikov} to solve the model \eref{eq.H}.
In this scheme one uses a double-time Green's
function $\gf{\hat{A}(t)}{\hat{B}(t')}$ ($\hat{A}$ and $\hat{B}$ represent two
arbitrary quantum operators) which satisfies the following equation of motion:
\begin{eqnarray}
\rmi\hbar\frac{\mathrm{d}}{\mathrm{d}t}\gf{\hat{A}&(t)}{\hat{B}(t')}=\nonumber\\
&\delta(t-t')\ave{[\hat{A}(t),\hat{B}(t)]}+\gf{[\hat{A}(t),\hat{H}]}{\hat{B}(t')}.
\label{eq.GFeom}
\end{eqnarray}
This approach proves successful for various Heisenberg spin 
models \cite{gf,Tyablikov,gf-spinS,CallenExp,lbg1,lbg3}.
In  \onlinecite{CallenExp}, there is a detailed but somewhat long derivation about the 
Green's function method for Heisenberg spin model. 
Accordingly, we sum up the derivation in  \onlinecite{CallenExp} and get the point
that the process of using Green's functions to solve the average spin of Heisenberg model can
be simplified into three steps: a) construct Green's functions
 $\gf{\hat{S}^+_{\mybf{i}}}{\hat{S}^-_{\mybf{j}}}$
and their equations of motion via \eref{eq.GFeom} and then
 use Tyablikov cutoff approximation \eref{eq.cutoff}
to decouple the equations of motion\cite{Tyablikov}; 
\begin{equation}
 \gf{\hat{S}^z_{\mybf{i}}\hat{S}^+_{\mybf{j}}}{\hat{B}} \longrightarrow \hspace{-1.6em}^{\mybf{i}\ne\mybf{j}} ~ \ave{\hat{S}^z}
    \gf{\hat{S}^+_{\mybf{j}}}{\hat{B}}
\label{eq.cutoff}
\end{equation}
b) use spectrum theorem to express the correlation function $\ave{\hat{S}^-\hat{S}^+}$ in term of
the average spin $z$-component $\ave{\hat{S}^z}$ and then get the $\Phi(\ave{\hat{S}^z})$ function
\begin{equation}
 \Phi(\ave{\hat{S}^z})=\frac12\ave{\hat{S}^-\hat{S}^+}/\ave{\hat{S}^z}~;
\label{eq.Phi}
\end{equation}
c) use the Callen Expression \eref{eq.Callen}\cite{CallenExp} to evaluate $\ave{\hat{S}^z}$
self-consistently
\begin{equation}
\ave{\hat{S}^z} =
\dfrac{(S-\Phi)(1+\Phi)^{2S+1}+(S+1+\Phi)\Phi^{2S+1}}{(1+\Phi)^{2S+1}-\Phi^{2S+1}}~.
\label{eq.Callen}
\end{equation}


According to the three steps given above, for our spin system we construct double-time
spin Green's functions between spin operators at two positions
$\mybf{i}$ and $\mybf{j}$ in SL1
\begin{equation}
\widetilde{G}_{\mybf{ij}}^{\sss{(11)}}(t,t')=\gf{\hat{S}_{1\mybf{i}}^+(t)}{\hat{S}_{1\mybf{j}}^-(t')}
\label{eq.G11}
\end{equation}
and Green's functions between spin operators at two positions
$\mybf{i}'$ in SL2 and $\mybf{j}'$ in SL1
\begin{equation}
\widetilde{G}_{\mybf{i}'\mybf{j}'}^{\sss{(21)}}(t,t')=
      \gf{\mySp{i\mbox{\unboldmath$\sss{'}$}}{\!\!-}(t)}{\hat{S}_{1\mybf{j}'}^-(t')}~.
\label{eq.G21}
\end{equation}
\Eref{eq.G11} can be expressed as Fourier expansion
\begin{equation}
\widetilde{G}^{\sss{(11)}}_{\mybf{ij}}(t,t')=\frac1{2\pi\hbar}\int
G^{\sss{(11)}}_{\mybf{ij}}(\omega)\rme^{-\rmi\omega(t-t')/\hbar}\mathrm{d}\omega,
\label{eq.FT}
\end{equation}
because the Hamiltonian \eref{eq.H} is time independent, and \eref{eq.G21}
is similar. Assuming that each spin has the same average $\ave{\hat{S}_1^z}$
for SL1 and $\ave{\hat{S'_2}{\!^z}}$ for SL2 and because of the AFM
symmetry plus the transformation \eref{eq.SS}, we have
$\ave{\hat{S}_1^z}=\ave{\hat{S'_2}{\!^z}}=\ave{\hat{S}^z}$.
Then using the Fourier transformation as shown in \eref{eq.FT},
making Tyablikov cutoff approximation to decouple the equations of motion
and another Fourier transformation from lattice sites real space
to $\mybf{k}$ space, we can have
\begin{equation}
1+g_{\mybf{k}}^{\sss{(11)}} \bigg [ J_{1b}\rho_{_1}(\mybf{k}) - \frac{\omega}{2\ave{\hat{S}^z}} \bigg ]
  +J_{1b}g_{\mybf{k}}^{\sss{(21)}} \rho_{_2}(\mybf{k})=0~
\label{eq.eom1}
\end{equation}
and
\begin{equation}
g_{\mybf{k}}^{\sss{(11)}} J_{1b}\rho_{_2}(\mybf{k})
  +g_{\mybf{k}}^{\sss{(21)}} \bigg[ J_{1b}\rho_{_1}(\mybf{k}) + \frac{\omega}{2\ave{\hat{S}^z}} \bigg ]=0~,
\label{eq.eom2}
\end{equation}
in which
\begin{equation}
\eqalign{
g_{\mybf{k}}^{\sss{(11)}}&=\sum_{\mybf{r}} G_{\mybf{i},\mybfa{i}{r}}^{\sss{(11)}} (\omega) \mathrm{e}^{-\rmi\mybf{k}\cdot\mybf{r}}\\
g_{\mybf{k}}^{\sss{(21)}}&=\sum_{\mybf{r}} G_{\mybf{i},\mybfa{i}{r}}^{\sss{(21)}} (\omega) \mathrm{e}^{-\rmi\mybf{k}\cdot\mybf{r}}~,
}
\label{eq.gk}
\end{equation}
\begin{equation}
\eqalign{
\rho_{_1}(\mybf{k}) &= -p(1-\cos\mybf{k}\!\cdot\!\bm{a})+1+2q+r\\
\rho_{_2}(\mybf{k}) &= (1+2q\cos\mybf{k}\!\cdot\!\bm{a})\cos\mybf{k}\!\cdot\!\bm{b}+r\cos\mybf{k}\!\cdot\!\bm{c}~,
}
\label{eq.rho}
\end{equation}
and
\begin{equation}
p\equiv\frac{J_{1a}}{J_{1b}}~,~~ q\equiv\frac{J_{2}}{J_{1b}}~,~~ r\equiv\frac{J_{c}}{J_{1b}}~.
\label{eq.pqr}
\end{equation}
We should point out that: a) the wave vector $\mybf{k}$ we used here is based
on the whole lattice sites (SL1+SL2), so $\mybf{r}$ in summations of \eref{eq.gk}
runs over all sites in the whole lattice; b) for homogeneous system
$G_{\mybf{i},\mybfa{i}{r}}^{\sss{(11)}}$ is only a function of relative position
$\mybf{r}$ and independent of $\mybf{i}$
(as a result $G_{\mybf{ii}}^{\sss{(11)}}=G^{\sss{(11)}}$ which is used below);
$G_{\mybf{i},\mybfa{i}{r}}^{\sss{(21)}}$ is analogous.

From \eref{eq.eom1} and \eref{eq.eom2} we derive
\begin{equation}
g_{\mybf{k}}^{\sss{(11)}}= \frac{\ave{\hat{S}^z}}{\sqrt{\rho_{_1}^2\!-\!\rho_{_2}^2}}
  \bigg[ \frac{\rho_{_1}\!+\!\sqrt{\rho_{_1}^2\!-\!\rho_{_2}^2}}{\omega-E(\mybf{k})}
        -\frac{\rho_{_1}\!-\!\sqrt{\rho_{_1}^2\!-\!\rho_{_2}^2}}{\omega+E(\mybf{k})}  \bigg],
\label{eq.gk11}
\end{equation}
where $E({\mybf{k}})$ is the spin excitation spectrum defined by
\begin{equation}
E(\mybf{k})=2J_{1b}\ave{\hat{S}^z} \sqrt{\rho_{_1}^2(\mybf{k})-\rho_{_2}^2(\mybf{k})}~.
\label{eq.Ek}
\end{equation}
And from $g_{\mybf{k}}^{\sss{(11)}}$ we get  $G_{\mybf{ij}}^{\sss{(11)}}$
\begin{equation}
G_{\mybf{ij}}^{\sss{(11)}}(\omega)=\frac1N \sum_{\mybf{k}\in\text{BZ}}
    g_{\mybf{k}}^{\sss{(11)}} \mathrm{e}^{\rmi\mybf{k}\cdot(\mybf{i}-\mybf{j})}~,
\label{eq.sumgk}
\end{equation}
in which $N$ is the total number of spins in SL1 and SL2, and BZ denotes the
first Brillouin zone (there are $N$ $\mybf{k}$-points in BZ).
Using spectrum theorem and letting $\mybf{j}=\mybf{i}$ we get  the correlation
function $\ave{\hat{S}_1^-\hat{S}_1^+}$ as follow
 \begin{equation}
 \eqalign{
 \ave{\hat{S}_1^-\hat{S}_1^+} &= -\frac1{\pi} \int_{-\infty}^{\infty}
        \dfrac{\mathrm{Im} [ G^{\sss{(11)}}(\omega+i0^+) ] }{\mathrm{e}^{\beta \omega}-1} \mathrm{d}\omega \\
     &= \frac{\ave{\hat{S}_1^z}}{N} \sum_{\mybf{k} \in \text{BZ}}
        \bigg[ \frac{\rho_{_1}}{\sqrt{\rho_{_1}^2 - \rho_{_2}^2}} \coth\frac{\beta E(\mybf{k})}{2} - 1 \bigg]~,
 }
 \label{eq.aveSS}
 \end{equation}
where $\beta=1/(k_{_\text{B}}T)$, $T$ is temperature, and $k_{_\text{B}}$ is the Boltzmann constant.
Then using \eref{eq.Phi} we get the $\Phi$ function:
\begin{equation}
\Phi(\ave{\hat{S}^z})=\frac1{2N}\sum_{\mybf{k}\in\text{BZ}}
  \bigg[ \frac{\rho_{_1}}{\sqrt{\rho_{_1}^2-\rho_{_2}^2}} \coth\frac{\beta E(\mybf{k})}{2} - 1 \bigg]~.
\label{eq.Phi2}
\end{equation}

Now, the average spin $z$-component $\ave{\hat{S}^z}$ can be
obtained easily by self-consistently solving
 \eref{eq.Phi2} and \eref{eq.Callen}. A special case is that when the temperature
$T=0$, $\coth\frac{\beta E(\mybf{k})}{2}\rightarrow 1$ and we have
\begin{equation}
\Phi_0\equiv\Phi|_{T=0}= \frac1{2N}\sum_{\mybf{k}\in\text{BZ}}
     \bigg[ \frac{\rho_{_1}}{\sqrt{\rho_{_1}^2-\rho_{_2}^2}} - 1 \bigg]~.
\label{eq.Phi0}
\end{equation}
At this time, $\Phi_0$ is no longer dependent on $\ave{\hat{S}^z}$  and
the zero-temperature average spin $z$-component $\avez{\hat{S}^z}$ can be
obtained directly by insert \eref{eq.Phi0} into \eref{eq.Callen}.

While temperature approaches to $\TN$, $\ave{\hat{S}^z}$  approaches to
zero and further $E(\mybf{k})\rightarrow 0$ and $\Phi\rightarrow\infty$.
Expanding \eref{eq.Phi2} and \eref{eq.Callen}, we derive
\begin{equation}
\ave{\hat{S}^z} \propto  \sqrt{ 1-\frac{T}{\TN} }~,
\label{eq.aveS2}
\end{equation}
where $\TN$ is defined by
\begin{equation}
\TN=\frac{2J_{1b}S(S+1)}{3\Gamma k_{_\text{B}}}
\label{eq.TN}
\end{equation}
and $\Gamma=\frac1N\sum_{\mybf{k}} [\rho_{_1}/(\rho_{_1}^2-\rho_{_2}^2)]$.

\section{\label{sec.result}Numerical Results and discussions}
\begin{figure}[!htb]
\begin{center}
\includegraphics[width=8cm]{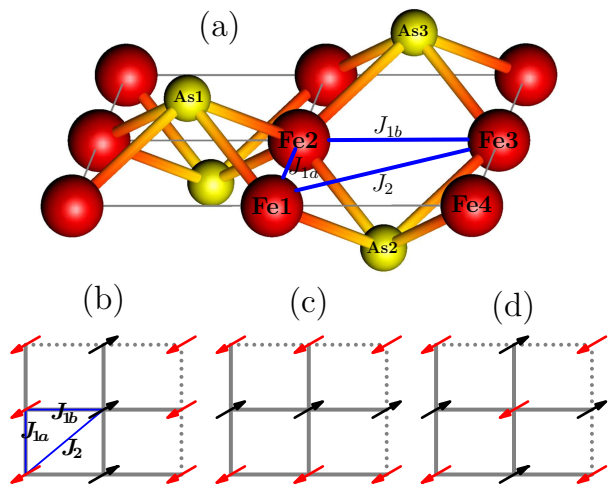}
\end{center}
\caption{(a) Structure of FeAs layer and scheme for the exchange interactions
mediated by Fe-As-Fe paths: Fe1-As1-Fe2 and Fe1-As2-Fe2 for $J_{1a}$, Fe2-As2-Fe3 and Fe2-As3-Fe3
for $J_{1b}$, and Fe1-As2-Fe3 for $J_2$. Different AFM configurations are shown: (b) stripe-AFM
along $\bm{a}$ direction, (c) stripe-AFM along $\bm{b}$ direction, and (d) checkerboard AFM.
Energies of the three AFM configurations are given in \eref{eq.emin}.
}
\label{fig:scheme}
\end{figure}

About the coupling interactions $J_{1a}$, $J_{1b}$, $J_2$ and $J_c$ in
FeAs-based pnictides, there is no consensus on their magnitudes and many authors
only consider two or three of them\cite{550K,Yildirim,superex,Ba2,sw-SrFe2As2,Ca1,lgq}.
Yildirim's first-principles results show that $J_1\sim J_2$\cite{Yildirim}.
However, we prefer the opinion that $J_{1a}$, $J_{1b}$ and $J_2$ originate from
AFM superexchange through As atoms\cite{superex1,superex2,superex3}. From the
viewpoint of the structure of FeAs layers, both $J_{1a}$ and
$J_{1b}$ are mediated by two Fe-As-Fe paths (\fref{fig:scheme}(a)) and there should be $J_{1a} \ne J_{1b}$
but  $J_{1a}\sim J_{1b}$ due to the small structure change from tetragonal to
orthorhombic symmetry. $J_2$ is mediated by only one Fe-As-Fe path (\fref{fig:scheme}(a)) and there should
be $2J_2 \sim J_{1a}$. From the viewpoint of classical favorable energy to
form stripe-like AFM patterns along $\bm{a}$ direction as shown in \fref{fig:scheme}(b)
(see \fref{fig:scheme}(b)-(d) and \eref{eq.emin}),
\begin{equation}
\left\{
\eqalign{
E_{\text{(b)}}&=\phantom{+}4J_{1a}S^2-4J_{1b}S^2-8J_2S^2\\
E_{\text{(c)}}&=-4J_{1a}S^2+4J_{1b}S^2-8J_2S^2\\
E_{\text{(d)}}&=-4J_{1a}S^2-4J_{1b}S^2+8J_2S^2~\\
E_{\text{(b)}}&<E_{\text{(c)}}~ \Longrightarrow~ J_{1a}<J_{1b}\\
E_{\text{(b)}}&<E_{\text{(d)}}~ \Longrightarrow~ 2J_{2}>J_{1a}
}
\right.
\label{eq.emin}
\end{equation}
there should be $J_{1a} < J_{1b}$ and $2J_2 > J_{1a}$,
which in fact are just the conditions that fulfill $\rho_{_1}^2-\rho_{_2}^2 \geqslant 0$
to make $E(\mybf{k})$ in \eref{eq.Ek} meaningful.
As for $J_c$, it's a very weak long-range AFM interaction with  a nowadays
unclear origin other than superexchange.  Therefore, in terms of $p,~q,~r$
in \eref{eq.pqr} we confine the coupling interactions as follow:
\begin{equation}
\left \{
\eqalign{
& 1 \geqslant p \geqslant 1-\delta_p\\
& p/2 <  q  \leqslant p/2+\delta_q~,\\
& 0 <r \leqslant \delta_r
} \right .
\label{eq.pqrrange}
\end{equation}
where $\delta_p$, $\delta_q$, and $\delta_r$ are the confine parameters for
$p$, $q$, and $r$ respectively and they fulfill $0<\delta_p,\delta_q,\delta_r\ll 1$.
Lhs of \eref{eq.pqrrange}
represents the necessary conditions to form striped AFM ordering along $\bm{a}$ direction,
and rhs the conditions to limit $p$, $q$ and $r$ within small regions.

\begin{figure}[!tbp]
\begin{center}
\includegraphics[width=8cm]{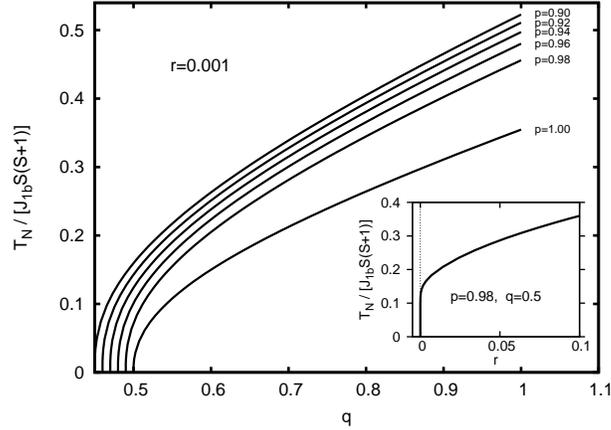}
\end{center}
\caption{The reduced AFM transition temperatures
($\TN/[J_{1b}S(S+1)]$) as functions of $q$ for $r=0.001$ and
$p=0.90,0.92,\cdots,1.0$ (from top to bottom). The inset shows an
$r$ dependence of $\TN$ for $p=0.98$ and $q=0.5$.} \label{fig:TN}
\end{figure}

From \fref{fig:TN} we can see that $\TN$ increases as $q$ and $r$ increase
but decreases as $p$ increases. It's easy to understand. NNN spins and inter-layer
NN spins all align antiparallel, so bigger AFM coupling interactions will lower
the system's energy, stabilize the AFM configuration and hence enhance $\TN$;
on the contrary, spins along $\bm{a}$ direction align parallel but have AFM
interactions, therefore bigger $p$ will increase the system's energy, destabilize
the AFM configuration and hence decrease $\TN$. It's notable that
while $q\rightarrow p/2$ or $r\rightarrow 0$, then $\TN\rightarrow 0$,
that is to say, both the existence of an AFM inter-layer interaction $J_c$
and the condition $2J_{2}>J_{1a}$ are  essential to form striped AFM ordering.
In fact, when $r=0$ this system becomes two-dimensional, and the result of $\TN=0$
in two dimensions is analogous to the Mermin-Wagner theorem for isotropic interactions\cite{MWT}.
We also see that the critical condition $p=1$ doesn't lead to $\TN\rightarrow 0$,
which manifests $J_{1a}=J_{1b}$ isn't a fatal factor to kill $\TN$ but only
a critical value to separate the two cases shown in \fref{fig:scheme}(b) and \ref{fig:scheme}(c).

\begin{figure}[!htbp]
\begin{center}
\includegraphics[width=8cm]{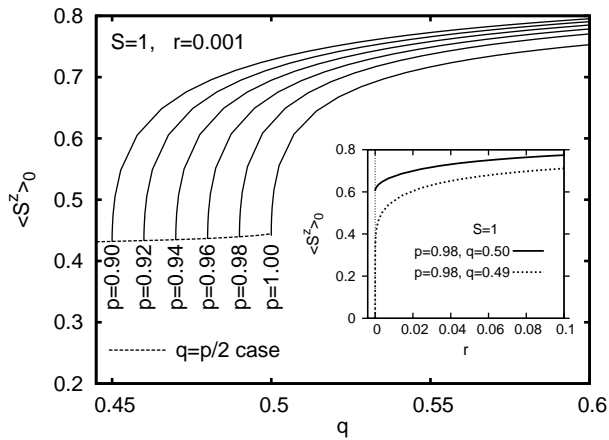}
\end{center}
\caption{The zero-temperature average spins $\avez{\hat{S}^z}$ as
functions of $q$ for $r=0.001$ and different $p$ values:
$0.90,0.92,\cdots,1.0$. The lower limits of $\avez{\hat{S}^z}$ are
shown by the dotted line. The inset shows $\avez{\hat{S}^z}\sim r$
curves for $(p,q)=(0.98,0.50)$ (solid) and $(0.98,0.49)$ (dotted)
respectively. } \label{fig:S0}
\end{figure}

Magnetic moment per Fe of LaFeAsO at low temperature is reported as
0.36$\mu_{\text{B}}$\cite{nat453_899} or 0.25$\mu_{\text{B}}$\cite{prl101_077005} experimentally,
both of which are very small compared with the first-principles values
$2.2\sim 2.4\mu_{\text{B}}$/Fe\cite{semimetal,Yildirim2}.
Accordingly, we choose $S=1$ in our model (assuming the Land\'e $g$-factor equals 2).
\Fref{fig:S0} shows the zero-temperature average spin  $\avez{\hat{S}^z}$.
Similar to $\TN$, $\avez{\hat{S}^z}$ is also an increasing function of both $q$ and $r$
but a decreasing function of $p$. However,  $\avez{\hat{S}^z}\rightarrow0$ only when both
$r\rightarrow0$ and $q\rightarrow p/2$.  When only $r\rightarrow0$ or $q\rightarrow p/2$ is
met,  $\avez{\hat{S}^z}$ approaches to a minimum but not zero, while at the same time $\TN\rightarrow0$.

\begin{figure}[!tbp]
\begin{center}
\includegraphics[width=8cm]{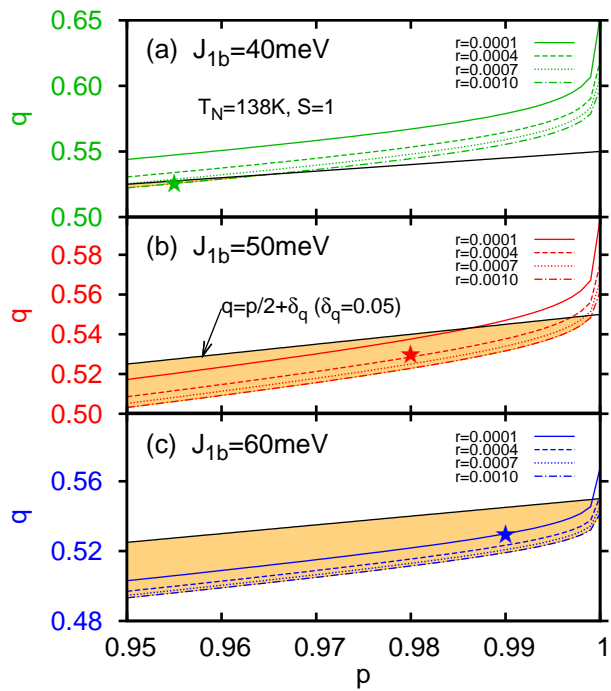}
\end{center}
\caption{The parameter regions (orange or gray)
satisfying \eref{eq.pqrrange} with $\TN=138$K for $J_{1b}=40$meV
(a), $J_{1b}=50$meV (b), and  $J_{1b}=60$meV (c), where we use
$\delta_p=0.05, \delta_q=0.05$, and $\delta_r=0.001$. The star
shows the value $(p,q,r)$=(0.955, 0.526, 0.0009) in (a), (0.98,
0.53, 0.0003) in (b), or (0.99, 0.53, 0.0001) in (c). }
\label{fig:pqr}
\end{figure}

There are four $J$'s in our model. What are their values?
Then let us have a look at what can they be under condition $\TN=138$K.
\Fref{fig:pqr} shows the regions available for $p$, $q$ and $r$ under conditions
\eref{eq.pqrrange} with $\delta_p=0.05,\delta_q=0.05$ and $\delta_r=0.001$
in orange (gray) colour for  $J_{1b}=40$meV, 50meV, and 60meV
respectively. The smaller  $J_{1b}$, the smaller the parameter 
region available. Hence, for given  $\delta_p$, $\delta_q$ and $\delta_r$,
there is a lower limit for $J_{1b}$  to fulfill a given $\TN$.
This lower limit, written as $J_{1b}^{\text{min}}$, is given in \tref{tab.J1bmin},
which shows that the smaller each of $\delta_p$, $\delta_q$ and $\delta_r$ is,
the bigger  $J_{1b}^{\text{min}}$ is.
However, although no upper limit for $J_{1b}$ is given out,
$J_{1b}$ can not be infinitely great.
In fact, first-principles results show that $J_{1b}\sim 50$meV\cite{550K,superex}.

Here we didn't use the experimental data for low-temperature magnetic moment
per Fe atom, which amounts to $\avez{\hat{S}^z}=0.13\sim0.18$, to determine the $J$'s,
because it's too small. If $\avez{\hat{S}^z}=0.18$ is met, there have to be
$\delta_r<10^{-6}$ even if $q=p/2$ is taken to minimize $\avez{\hat{S}^z}$
(see inset of \fref{fig:S0}). Such tiny $r$ definitely cannot fall within
the orange (gray) area in \fref{fig:pqr} with a reasonable $J_{1b}$ to fulfill
$\TN=138$K. That is to say, although the many-body AFM fluctuations
substantially reduce the low-temperature magnetic moment per Fe, yet they
cannot be in full charge
of the very small low-temperature magnetic moment which indeed is also ascribed
to spin orbit and $p$-$d$ hybridization etc\cite{prl101_126401}.

\begin{table}[!htbp] 
\caption{\label{tab.J1bmin} $J_{1b}^{\text{min}}$ for given
 $\delta_p$, $\delta_q$ and $\delta_r$.  }
\begin{indented}\item[]
\newcommand{\myph}{$\phantom{0}$}
\begin{tabular}{cccc}\br
 $\hspace{5mm}\delta_q\hspace{5mm}$ & $\delta_r$ & $\delta_p$ & $J_{1b}^{\text{min}}$/meV \\\mr
 0.01\myph  & 0.0001     & 0.01       &     104.4     \\
            &            & 0.1\myph   & \myph89.8     \\
            & 0.001\myph & 0.01       & \myph87.1     \\
            &            & 0.1\myph   & \myph72.6     \\\mr
 0.05\myph  & 0.0001     & 0.01       & \myph50.9     \\
            &            & 0.1\myph   & \myph44.6     \\
            & 0.001\myph & 0.01       & \myph43.6     \\
            &            & 0.1\myph   & \myph37.2     \\
\br
\end{tabular}
\end{indented}
\end{table}

\begin{figure}[!tbp]
\begin{center}
\includegraphics[width=8cm]{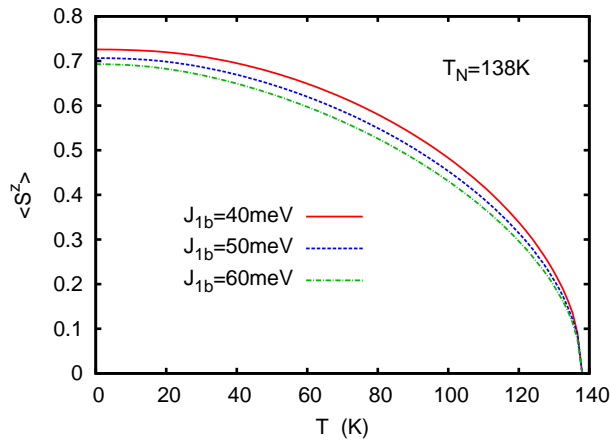}
\end{center}
\caption{The average spins $\ave{\hat{S}^z}$ as
functions of the temperature $T$ for $J_{1b}$=40meV, 50meV, and
60meV. The corresponding  $(p,q,r)$ parameters are (0.955, 0.526,
0.0009), (0.98, 0.53, 0.0003), and (0.99, 0.53, 0.0001),
respectively.} \label{fig:ST}
\end{figure}

We choose three sets of $J$'s from the regions available in
\fref{fig:pqr} for estimation (the three stars): $(p,q,r)$=(0.955, 0.526,
0.0009), (0.98, 0.53, 0.0003), and (0.99, 0.53, 0.0001) for $J_{1b}$=40meV,
50meV, and 60meV respectively. In terms of $J_{1b}$, $J_{1a}$, $J_2$ and $J_c$,
we can refer to them as $J_{1b}=50\pm10$meV, $J_{1a}=49\pm10$meV,
$J_2=26\pm5$meV, and $J_c=0.020\pm0.015$meV. The average spin $\ave{\hat{S}^z}$
vs temperature $T$ curves with the three sets of parameters given above are
shown in \fref{fig:ST}.

We take LaFeAsO for example here, however our calculations are not restricted
to LaFeAsO, because nearly all $Ln$FeAsO has similar even the same transition
temperatures (see \tref{tab:LnOFeAs}).
\begin{table}[!htbp]
\caption{ \label{tab:LnOFeAs} AFM and structure transition
temperatures $\TN$ and $T_\text{S}$ of $Ln$FeAsO. }
\begin{indented}\item[]
\begin{tabular}{cccccc}\br
  $Ln$            &   La   &   Ce  &  Pr  &  Nd  &  Sm  \\\mr
$\TN$(K)          &   138  &   140 &  127 &  141 &  140 \\
$T_\text{S}$(K)  &   156  &   155 &  153 &   -  &   -  \\
  Ref            &\onlinecite{prl101_077005}&\onlinecite{Ce1}& \onlinecite{Pr1} &\onlinecite{Nd1}&\onlinecite{Sm4}\\
\br
\end{tabular}
\end{indented}
\end{table}
It seems that $\TN$ doesn't vary much with different $Ln$. This is different from
$A$Fe$_2$As$_2$, whose $\TN$ varies with $A$: $\TN$=172.5K\cite{Ca3} for $A$=Ca;
$\TN$=198K\cite{Sr2}, 205K\cite{Sr4} or 220K\cite{Sr5} for $A$=Sr and
$\TN$=140K\cite{Ba1} or 135K\cite{Ba4} for $A$=Ba.
Although there are some differences of structure between  $Ln$FeAsO and $A$Fe$_2$As$_2$,
we believe that our model works well for both of them, because they both have layered
structures with  stripe-like AFM order formed by Fe atoms. Indeed, this can also be extended to
Fe(SeTe)\cite{FeSe1,FeTeSe2}, which also have nearly the same
properties and can be superconducting under certain conditions.

\section{\label{sec.con}Conclusion}

In summary, we use a Green's function method to study the striped
AFM order formed by Fe atoms in $Ln$FeAsO which is the
representative of the parent compounds of recently discovered
Fe-based superconductors. We take LaFeAsO for example to analyze
the AFM transition temperature $\TN$ and zero-temperature average
spin $\avez{\hat{S}^z}$, and show that both $\TN$ and
$\avez{\hat{S}^z}$ are increasing functions of $J_{1b}$, $J_2$ and
$J_c$ but decreasing functions of $J_{1a}$. By using $\TN=138$K,
we make a reasonable estimation of the coupling interactions,
$J_{1b}=50\pm10$meV, $J_{1a}=49\pm10$meV, $J_2=26\pm5$meV, and
$J_c=0.020\pm0.015$meV. Average spin $\ave{\hat{S}^z}$ is
determined in the same way from zero temperature to $\TN$ and
$\TN$ is expressed analytically. Our results also show that a
non-zero AFM inter-layer coupling $J_c$ is essential to the
existence of a non-zero $\TN$ and that the AFM fluctuations
substantially reduce the low-temperature magnetic moment towards
the small experimental value. Although our results cannot
determine the relations between structure and AFM transitions, we
believe that the AFM transition is likely caused by the structure
transition because most of experimental results show $\TN\leqslant
T_\text{S}$ for FeAs-based pnictides. Our Green's function
approach of the striped AFM properties can be used to other
FeAs-based parent compounds.

\ack
This work is supported  by Nature Science Foundation of China
(Grant Nos. 10874232 and 10774180), by the Chinese Academy of
Sciences (Grant No. KJCX2.YW.W09-5), and by Chinese Department of
Science and Technology (Grant No. 2005CB623602).



\section*{References}

\providecommand{\newblock}{}

\end{document}